\pdfoutput=1
\documentclass[aps,onecolumn,superscriptaddress]{revtex4-2}
\usepackage{graphicx}
\usepackage{dcolumn}
\usepackage{bm}

\usepackage{amsmath}
\usepackage{amssymb}
\usepackage{braket}
\usepackage[caption=false]{subfig}
\usepackage[colorlinks=true]{hyperref}
\usepackage{microtype}
\usepackage{comment}

\usepackage{xcolor}

\usepackage{ulem} 

\usepackage{xcolor}

\usepackage{etoolbox}
\patchcmd{\section}
  {\centering}
  {\raggedright}
  {}
  {}
\patchcmd{\subsection}
  {\centering}
  {\raggedright}
  {}
  {}

\makeatletter
\renewcommand{\fnum@figure}{\textbf{Fig.~\thefigure}}
\renewcommand{\@caption@fignum@sep}{\ \textbf{\textbar}\ }
\makeatother

\begin{document}

\title{Uncooled low-noise thin-film optomechanical resonator for thermal sensing on lithium niobate}

\author{Yue Yu}
\thanks{These authors contributed equally to this work.}
\affiliation{Department of Electrical Engineering and Computer Sciences, University of California, Berkeley, Berkeley, CA 94720, USA}
\affiliation{Ming Hsieh Department of Electrical and Computer Engineering, University of Southern California, Los Angeles, CA, 90089, USA}
\affiliation{Materials Sciences Division, Lawrence Berkeley National Laboratory, Berkeley, CA, USA}

\author{Ran Yin}
\thanks{These authors contributed equally to this work.}
\affiliation{Department of Electrical Engineering and Computer Sciences, University of California, Berkeley, Berkeley, CA 94720, USA}
\affiliation{Ming Hsieh Department of Electrical and Computer Engineering, University of Southern California, Los Angeles, CA, 90089, USA}

\author{Ian Anderson}
\affiliation{Chandra Family Department of Electrical and Computer Engineering, The University of Texas at Austin, TX, 78712, USA}

\author{Yinan Wang}
\affiliation{Chandra Family Department of Electrical and Computer Engineering, The University of Texas at Austin, TX, 78712, USA}

\author{Jack Kramer}
\affiliation{Chandra Family Department of Electrical and Computer Engineering, The University of Texas at Austin, TX, 78712, USA}

\author{Chun-Ho Lee}
\affiliation{Department of Electrical Engineering and Computer Sciences, University of California, Berkeley, Berkeley, CA 94720, USA}
\affiliation{Ming Hsieh Department of Electrical and Computer Engineering, University of Southern California, Los Angeles, CA, 90089, USA}

\author{Xinyi Ren}
\affiliation{Department of Electrical Engineering and Computer Sciences, University of California, Berkeley, Berkeley, CA 94720, USA}
\affiliation{Ming Hsieh Department of Electrical and Computer Engineering, University of Southern California, Los Angeles, CA, 90089, USA}

\author{Zaijun Chen}
\affiliation{Department of Electrical Engineering and Computer Sciences, University of California, Berkeley, Berkeley, CA 94720, USA}
\affiliation{Ming Hsieh Department of Electrical and Computer Engineering, University of Southern California, Los Angeles, CA, 90089, USA}

\author{Michelle Povinelli}
\affiliation{Ming Hsieh Department of Electrical and Computer Engineering, University of Southern California, Los Angeles, CA, 90089, USA}

\author{Dan Wasserman}
\affiliation{Chandra Family Department of Electrical and Computer Engineering, The University of Texas at Austin, TX, 78712, USA}

\author{Ruochen Lu}
\affiliation{Chandra Family Department of Electrical and Computer Engineering, The University of Texas at Austin, TX, 78712, USA}

\author{Mengjie Yu}
\affiliation{Department of Electrical Engineering and Computer Sciences, University of California, Berkeley, Berkeley, CA 94720, USA}
\affiliation{Ming Hsieh Department of Electrical and Computer Engineering, University of Southern California, Los Angeles, CA, 90089, USA}
\affiliation{Materials Sciences Division, Lawrence Berkeley National Laboratory, Berkeley, CA, USA}


\date{\today}

\begin{abstract} 
Optomechanical transduction harnesses the interaction between optical fields and mechanical motion to achieve sensitive measurement of weak mechanical quantities with inherently low noise. 
Lithium niobate combines low optical loss, strong piezoelectricity, high intrinsic $f\times Q_m$ factor, and low thermal conductivity, making it promising for exploring optomechanical platforms targeting thermal sensing applications. Here, we developed an integrated optomechanical platform on thin-film lithium niobate with precisely engineered optical, mechanical, and thermal fields within a compact $40~\mu\text{m}\times40~\mu\text{m}$ footprint. The platform integrates suspended microring resonators with ultrathin central membranes, reducing mechanical stiffness and effective mass while maintaining a high optical factor $Q_o$ of $10^6$ and mechanical quality factor $Q_m$ of 1117, which increases to $5.1\times10^{4}$ after oscillation. The design suppresses thermal dissipation into the silicon substrate and enhances thermal sensitivity, achieving a temperature coefficient of frequency of $-124$ ppm/K and a noise-equivalent power of 6.2 $\text{nW}/\sqrt{\text{Hz}}$ at 10 kHz at room temperature. This compact and scalable platform opens up new opportunities for high-sensitivity thermal sensing, supports heterogeneous integration with infrared absorbers for uncooled infrared detection, and enables fully integrated, all-optical on-chip readout, paving the way toward large-format, low-noise infrared sensing arrays.
\end{abstract}

\maketitle

\section*{Introduction}
Advances in nanofabrication have enabled the integration of diverse functionalities onto a single chip, yielding smaller, faster, more cost-effective, and energy-efficient platforms.
With a tightly confined field and reduced mode volume, integrated platforms facilitate efficient transduction across different physical domains. Notably, cavity optomechanics leverages resonant optical confinement to enhance the interaction with mechanical motion, allowing the sensitive detection and control of weak displacements \cite{aspelmeyer2014cavity}. This approach underpins a broad range of applications, including high-precision sensing \cite{yu2016cavity,li2021cavity}, multiphysical transduction \cite{hill2012coherent,fang2016optical}, and quantum technologies \cite{barzanjeh2022optomechanics}. Reducing mechanical and optical losses is essential to these applications: Mechanical loss sets the thermal noise floor and constrains the coherence of mechanical motion, while optical loss limits light–mechanics interaction and readout sensitivity. On integrated platforms, optical loss can be minimized through fabrication optimization, whereas mechanical dissipation can be mitigated through several architectures, including optomechanical crystals with engineered bandgaps \cite{chan2012optimized,maccabe2020nano}, acoustic-wave-based devices \cite{balram2016coherent,diamandi2025optomechanical}, and microdisk/ring resonators with narrow supporting structures \cite{anetsberger2008ultralow,zhang2021optomechanical}. Among these, microdisk/ring resonators present a favorable balance between compact footprint, strong optomechanical coupling, electrode-free operation, and fabrication simplicity, making them especially attractive for practical implementations.

To date, silicon and silicon nitride have dominated optomechanical systems due to their mature fabrication ecosystem and CMOS compatibility, but they suffer from intrinsic material limitations such as weak piezoelectricity and optical nonlinearity. In contrast, lithium niobate (LN) provides a unique combination of low optical loss and large second-order optical nonlinearity \cite{weis1985lithium}, which has enabled various applications ranging from volt-level high-speed modulators to electro-optical frequency combs, broadband spectrometers, and photonic quantum networking and computing \cite{boes2023lithium}. Despite its strong piezoelectricity ($d_{33}\sim$ 6 pC/N) and intrinsic high $f\times Q_m$ product ($7.1\times 10^{13}$ \cite{bajak1981attenuation}), LN optomechanics remains less explored due to fabrication challenges. Recent advances demonstrated piezo-optomechanical racetrack resonators on thin film LN (TFLN) with optical quality factor ($Q_o$) of $2.1\times10^6$, mechanical quality factor ($Q_m$) of $5.5\times10^3$ at 200K near 5.2 GHz \cite{shen2020high}. 
Combined with its low thermal conductance (0.5 W/m·K) and high temperature coefficients of frequency ($|TCF| > 70$ ppm/K, which is over 2 times higher than other materials including silicon, silicon nitride, and aluminum nitride \cite{kim1969thermal,sui2025review}), LN holds great promise for developing sensitive uncooled optomechanical thermal sensors at room-temperature\cite{yu2025room,qian2025sensitive}.

Here, we demonstrate a room-temperature, low-noise optomechanical platform on TFLN by leveraging advanced nanotechnologies to simultaneously engineer optical, mechanical, and thermal domains. The platform integrates pixel arrays of suspended bowl-shaped optomechanical resonators, each pixel of $40~\mu\text{m}\times40~\mu\text{m}$, that combine a microring resonator with a thin-film membrane architecture inspired by two-dimensional material systems \cite{zheng2015hexagonal,blaikie2019fast,li2023terahertz}. The outer rim functions as a low-loss optical cavity, while the central region is trimmed into an ultrathin mechanical membrane of sub-100nm thick, substantially reducing the effective mass and stiffness to enhance thermal sensitivity without compromising strong optomechanical coupling. Each resonator is supported by four ultrathin anchors, which are carefully engineered to ensure mechanical stability while efficiently isolating the device from the underlying silicon thermal bath, thereby suppressing parasitic heat leakage and preserving high thermal sensitivity. Our optomechanical resonator, designed with a minimized bending radius of $20~\mu\text{m}$, shows an intrinsic $Q_o$ of $1.0\times 10^{6}$ after two-step etching and suspension, significantly exceeding previous demonstrations \cite{wang2014integrated,jiang2016chip,jiang2020efficient}. The fundamental radial mode at 83.7 MHz achieves a $Q_m$ of 1117 at room temperature and ambient pressure, increasing to $5.1\times10^{4}$ after oscillation. As a thermal sensor, the resonator is theoretically characterized to have a large thermal conductance of 33.5 $\mu\text{W/K}$ and a small thermal capacitance of 440 pJ/K, corresponding to a wide thermal bandwidth of 12.2 kHz enabled by its low thermal mass. Finally, the device demonstrates an intrinsic TCF of $-124$ ppm/K with a noise equivalent power (NEP) of 6.2 $\text{nW}/\sqrt{\text{Hz}}$ at 10 kHz. These results highlight our scalable and low-noise platform for exploring room-temperature optomechanics as the backbone for precision thermal sensors with frequency-shift modality and all-optical readout capability. Our resonators can be heterogeneously integrated with thin-film infrared (IR) absorbers \cite{guadagnini2025suspended}, accessing the new regimes of highly sensitive, large-scale, uncooled optomechanical IR detectors with revolutionary performance.

\begin{figure}
    \centering
    \includegraphics[width=\linewidth]{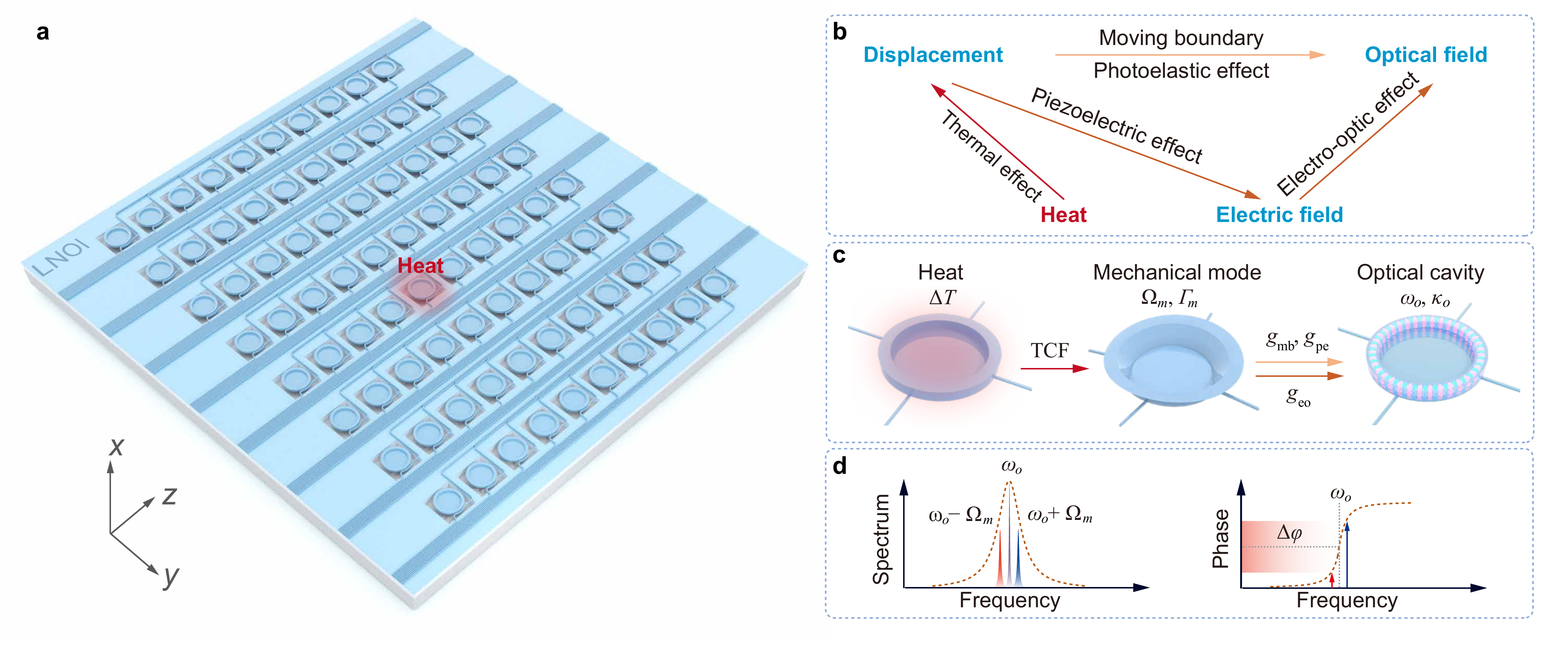}
    \caption{\textbf{Room-temperature optomechanical platform on TFLN.} \textbf{a.} Illustration of the bowl-shaped optomechanical resonator. \textbf{b.} Thermal sensing based on optomechanical transduction on LN. \textbf{c.} Transduction between thermal, mechanical, and optical fields with labeled parameters.  Mechanical frequency and damping rate $(\Omega_m, \Gamma_m)$, optical frequency and damping rate ($\omega_o, \kappa_o$), coupling rate induced by moving boundary, photoelastic, and electro-optic effects ($g_{mb}, g_{pe}, g_{eo}$). \textbf{d.} Transduction mechanism for the laser resonantly probing the cavity, where the mechanical motion causes the reflected field to be phase modulated.}
    \label{fig:1}
\end{figure}
\section*{Results}
Figure \ref{fig:1}{\color{red}a} illustrates our low-noise optomechanical platform fabricated on TFLN, where hundreds of pixels are integrated on a $1~\text{cm} \times 1~\text{cm}$ chip. On optomechanical platforms like silicon and silicon nitride, mechanical displacement can be directly coupled to the optical field through the moving boundary (MB) and photoelastic (PE) effect. Due to the strong electro-optic (EO) and piezoelectric effect of LN, an additional optomechanical coupling path arises, in which mechanical displacement introduces internal electrical field through piezoelectric effect that can further modulate the optical field through the EO effect (Fig. 1b). Heat accompanied by a temperature change $\Delta T$ is transduced to a mechanical frequency $\Omega_m$ shift via the thermomechanical effect. This frequency shift is subsequently read-out in the optical domain with a total single-photon optomechanical coupling rate $g_0 = g_{0,MB} + g_{0,PE} + g_{0,EO}$, defined as the frequency shift in the optical mode due to the zero-point motion of the mechanical oscillator \cite{shao2019microwave} (Fig.~\ref{fig:1}{\color{red}c}). With a zero-detuning optical pump coupled into the resonator, the thermomechanical vibration strongly modulates the optical phase, which can be precisely measured using balanced homodyne detection at the output (Fig.~\ref{fig:1}{\color{red}d}).

\begin{figure}
    \centering
    \includegraphics[width=\linewidth]{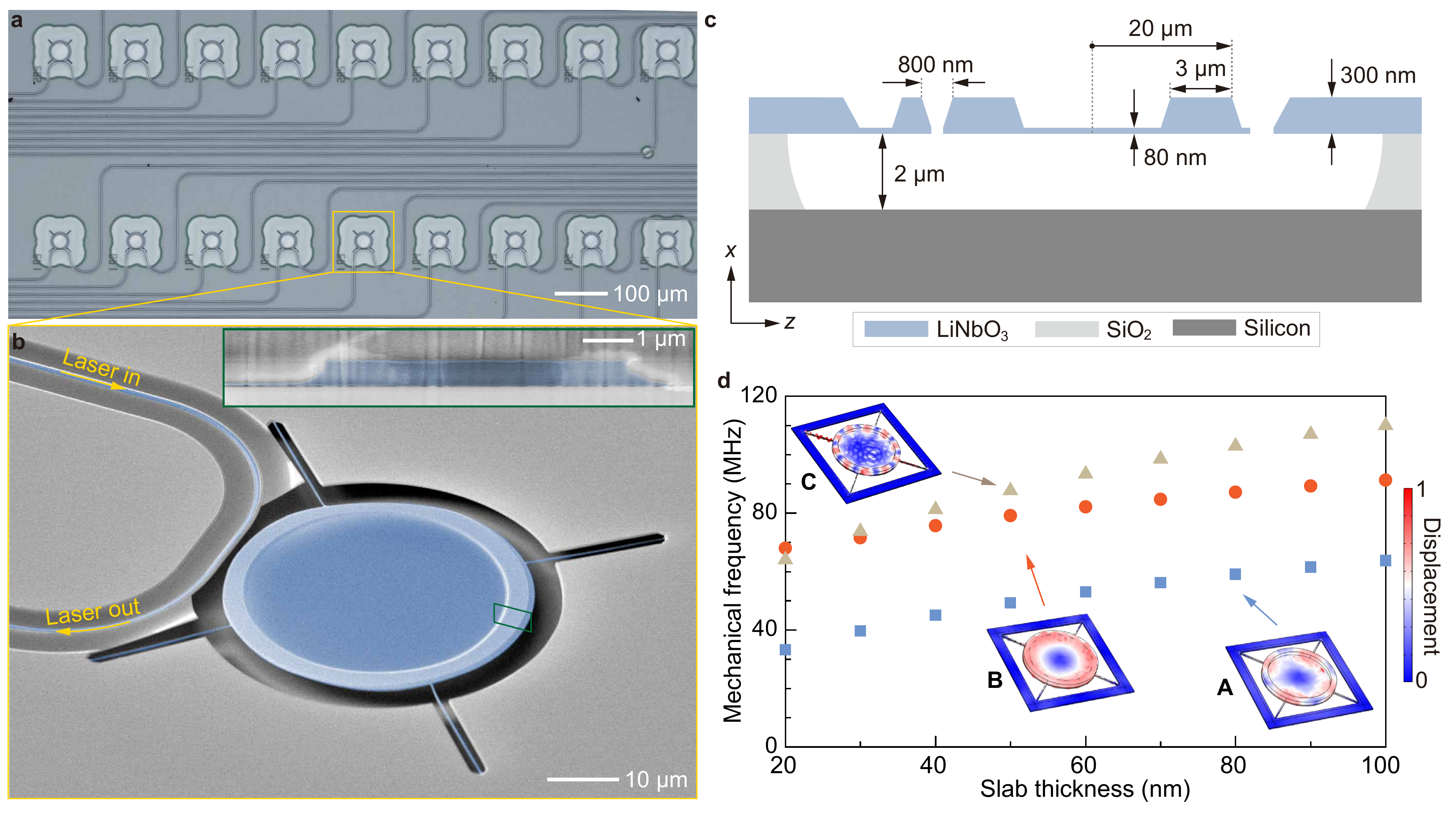}
    \caption{\textbf{Design and numerical simulation of bowl-shaped optomechanical resonators.} \textbf{a.} A microscope image of arrays of optomechanical pixels, demonstrating large scalability. \textbf{b.} SEM images of a fabricated resonator, the inset shows the cross-section. \textbf{c.} Side views of the device, with dimension labels. \textbf{d.} Simulated mechanical frequency of three modes at different slab thicknesses. The inserts represent the mechanical displacement profiles of corresponding modes.}
    \label{fig:2}
\end{figure}

To realize a low-noise optomechanical platform for future sensing applications, high-quality optical and mechanical resonances are essential to suppress thermal and readout noise and maintain coherence, along with a strong optomechanical coupling rate for efficient transduction between optical and mechanical degrees of freedom. We designed radial-mode bowl-shaped optomechanical resonators on 300-nm X-cut LN on insulator (Fig. \ref{fig:2}{\color{red}a}), which supports tightly confined optical modes circulating around its outer circumference, and strong mechanical vibrations enabled by a reduced inner membrane thickness (Fig. \ref{fig:2}{\color{red}b}). Figure \ref{fig:2}{\color{red}c} shows the cross-section of our device with labeled dimensions. The resonator has an outer and an inner radius of 20 and 17 $\mu \text m$ to balance the optical bending loss and the footprint of the device. Four anchors, each 400 nm wide and 20 $\mu \text m$ long, are fabricated on the slab layer and attached to the outer rim to provide mechanical support as well as minimize the perturbation of both optical and mechanical confinement. By balancing the device performance and fabrication challenges, we chose the anchor (and slab) thickness of 80 nm for fabrication, which included two-step e-beam lithography and etching, annealing, and suspension (see Methods for details). The small volume of the anchors is optimized to reduce the thermal conductance, therefore enhancing thermal sensitivity and making them suitable for on-chip bolometer applications. A bus waveguide is placed in close proximity to the resonator to couple light into and out of the resonator for detecting the thermomechanical vibration via an optomechanical transduction. Next, we simulate the frequencies of supported mechanical modes at different thicknesses of the inner membrane ($t_\text{slab}$). Figure \ref{fig:2}{\color{red}d} only plots the fundamental frequency of three mechanical modes (A,B,C) at near 80 MHz with their corresponding displacement profiles (embedded). Although high $f \times Q_m$ products can be achieved in higher-order mechanical modes \cite{kimble2009SiNfilms,Vengalattore2014membrane}, their proximity to numerous neighboring modes leads to elevated displacement background noise. For this reason, optimal low-noise operation is achieved by preferentially coupling to the fundamental mechanical mode of the resonator.

\begin{figure}
    \centering
    \includegraphics[width=\linewidth]{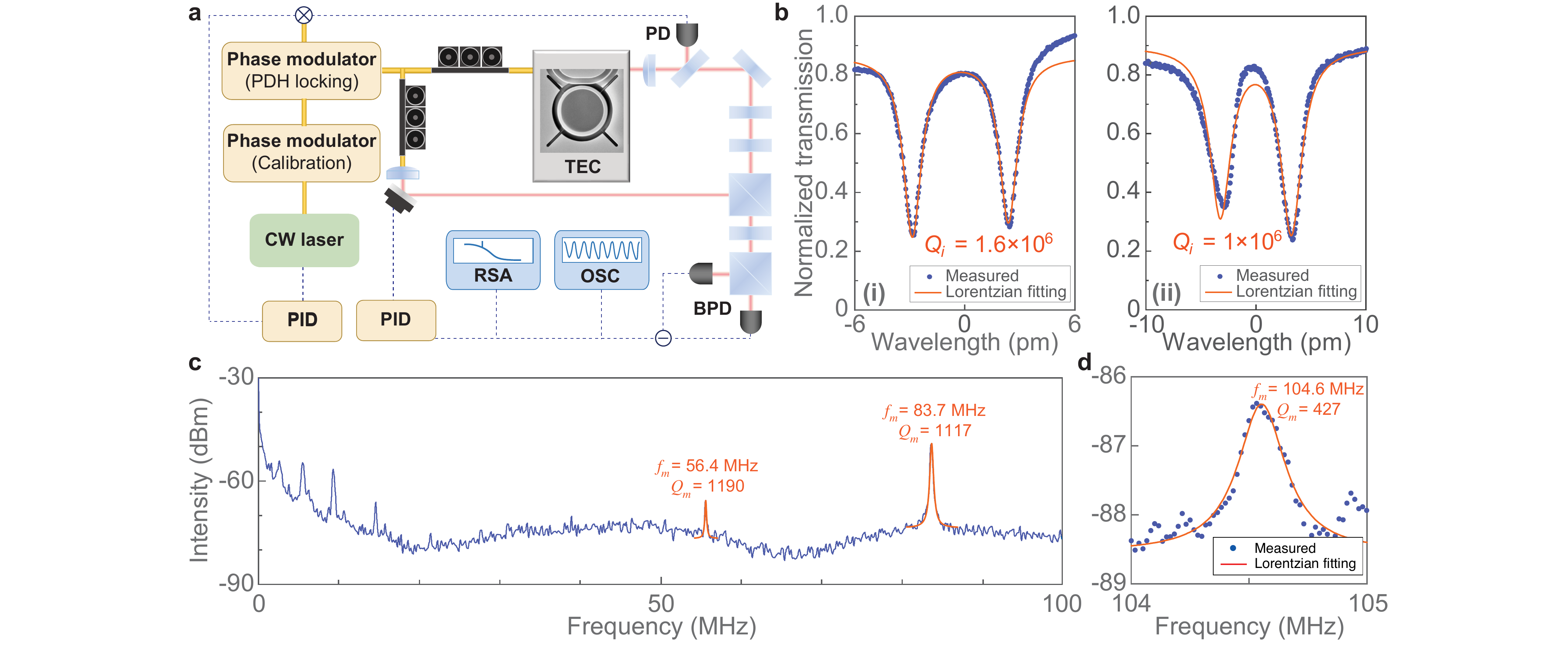}
    \caption{\textbf{Experimental setup and optomechanical characterization.}  \textbf{a.} Experimental setup, CW, continuous wave; PDH, Pound-Drever-Hall; PID, Proportional–Integral–Derivative controller; TEC, thermoelectric cooler; PD, photodetector; BPD, balanced photodetector; RSA, signal analyzer; OSC, oscilloscope. \textbf{b.} Measured optical resonance with Lorentzian-fitted optical $Q_i$ factors after first etching (i) and suspension (ii). \textbf{c.} Measured PSD through phase detection up to 100 MHz. \textbf{d.} Measured PSD through intensity detection near 104 MHz.}
    \label{fig:3}
\end{figure}
Figure \ref{fig:3}{\color{red}a} shows the experimental setup to measure the thermomechanical vibration using balanced homodyne detection with assistance of the Pound-Drever-Hall locking (see Methods for details). This scheme is used to suppress the laser-cavity detuning noise and the laser intensity noise for achieving a stable and precise optical readout. All measurements were conducted under room temperature and atmospheric pressure conditions, unless otherwise noted. To reduce the etching-induced optical loss, the anchors are positioned away from the outer rim of the resonator (see Supplementary Sec. 1). This strategy allows us to maintain low optical loss even after two-step etching, where the fabricated device shows an intrinsic quality factor $Q_o$ of $1.6 \times 10^6$ and $1 \times 10^6$ after the first etch (i) and the final suspension (ii), as shown in Fig.~\ref{fig:3}{\color{red}b}. The decreased resonance extinction ratio after suspension is mainly attributed to the vertical movement of the suspended resonator, which becomes more pronounced as the anchors become thinner, while most of the bus waveguide is still supported on the substrate. We then measure the displacement noise power spectral density (PSD) of mechanical modes using phase detection for frequency range below 100 MHz and intensity detection for frequency range above 100 MHz (Fig.~\ref{fig:3}{\color{red}c}, {\color{red}d}). With an on-chip power of $56~\mu\text{W}$ and a local oscillation power of $388~\mu\text{W}$, a maximum signal-to-noise ratio of 24 dB is achieved at 83.7 MHz with $Q_m$ of 1117, corresponding to the fundamental radial mode B shown in Fig.~\ref{fig:2}{\color{red}d}. 
\begin{figure}
    \centering
    \includegraphics[width=\linewidth]{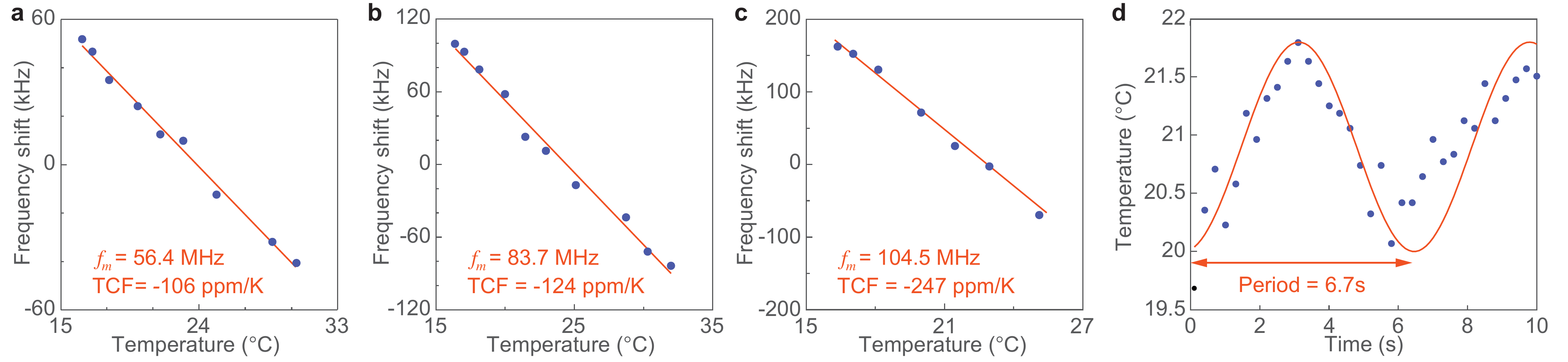}
    \caption{\textbf{Mechanical thermal sensitivity measurement.}  \textbf{a-c.} Measured TCFs of three mechanical modes are shown in Fig. 3 with a linear fitting curve. \textbf{d.} Reconstructed temperature signal from modulated mechanical frequency in the time domain, exhibiting periodic behavior consistent with the applied modulation frequency of 0.15 Hz.} 
    \label{fig:4}
\end{figure}

We then characterize the thermal sensitivity of the device by tuning the equilibrium temperature from 289 to 306 K using thermoelectric coolers (TECs) [see Supplementary Sec. 2]. Our resonator shows high TCF ranging from –106 to –263 ppm/K for different mechanical modes, as shown in Fig. 4a-c (also see Supplementary Sec.3). The fundamental mode located at 83.7 MHz has a TCF of –124 ppm/K, exceeding those of other integrated platforms ($|TCF|< 50$ ppm/K \cite{sui2025review}). To further demonstrate a dynamical response, we apply a sinusoidal wave voltage to the TEC and capture the mechanical frequency response in the time domain with an oscilloscope. With the measured TCF value, we can extract the temperature variation (Fig.~\ref{fig:4}{\color{red}d}), which shows a frequency of 0.15 Hz and a range of 2 °C. To further investigate the thermal bandwidth of our device, a faster modulation of the heat source is required, for example, by constructing a bolometer device where a modulated radiation signal within the absorption band is transduced to the temperature variation.

\begin{figure}
    \centering
    \includegraphics[width=\linewidth]{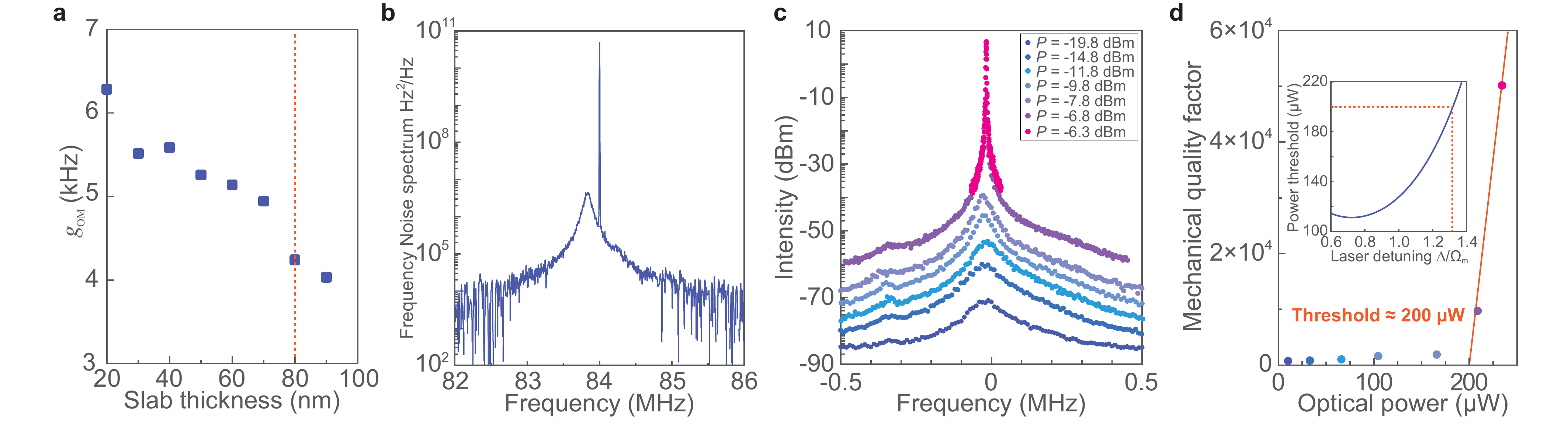}
    \caption{\textbf{Optomechanical coupling characterization.}  \textbf{a.} Simulated optomechanical coupling rate at different slab thickness, showing a $g_\text{OM}$ of 4.3 kHz at slab thickness of 80 nm. \textbf{b.} Applied calibration peak at 84 MHz, near the target frequency, used to extract the optomechanical coupling rate. \textbf{c.} Measured mechanical amplification as a function of increasing optical power, showing a drastic reduction in linewidth accompanied by a pronounced increase in oscillation intensity when exceeding the power threshold. \textbf{d.} Measured $Q_m$ at different optical powers with the inset showing the laser-detuning-dependent lasing threshold.}
    \label{fig:5}
\end{figure}

 We further characterize the single-photon optomechanical coupling rate $g_0$ of the fundamental radial mode by adding another phase modulation signal to the input light and generating the calibration sideband near the target frequency, as shown in Fig.~\ref{fig:5}{\color{red}b}. By comparing the calibration sideband power with the mechanical sideband power, we get the experimental $g_\text{OM}$ of 4.3 kHz, which is consistent with the simulation result shown in Fig. 5a at an 80-nm slab thickness. The performance of the optomechanical resonator can also be quantified by measuring the mechanical linewidth as a function of pumping power. Figure 5c plots measured RF spectra of cavity transmission at different pump powers. When the pump power is below the oscillation threshold, the mechanical spectrum closely resembles that of thermal mechanical vibration, as shown in Fig. 5b. As the pump power increases, the mechanical mode is driven into an oscillation state, and the peak value of the mechanical spectral intensity is enhanced by 80 dB. Simultaneously, the mechanical linewidth is drastically narrowed to 1.6 kHz, corresponding to an effective $Q_{m}$ of $5 \times 10^4$ (Fig. 5d). This behavior clearly demonstrates the mechanical lasing, and by fitting the above-threshold data with a linear curve, we extract a threshold power of 200 $\mu$W. The power threshold of the optomechanical lasing is given by \cite{jiang2012high}
\begin{equation}
P_\text{th}=\frac{m_\text{eff}\omega_{o}\Gamma_{i}\Gamma_{m}}{2g_\mathrm{OM}^2\Gamma\Delta}\left[(\Delta-\Omega_{m})^2+(\Gamma/2)^2\right]\left[(\Delta+\Omega_{m})^2+(\Gamma/2)^2\right]\frac{4\Delta^2+\Gamma^2}{4\Gamma_{i}\Gamma_{e}}
\end{equation}
where $P_\text{th}$ is the threshold optical power on to the chip, $m_\text{eff}$ is the effective motional mass, $\Gamma_{i}$, $\Gamma_{e}$, and $\Gamma$ are the intrinsic, external, and total photon decay rates, respectively. $\Gamma_{m}$ is the energy decay rate of the mechanical mode and $\Delta$ is the laser-cavity detuning. Using the measured resonator parameters, we can derive the theoretical power threshold of the device at different detunings, as shown in the insert of Fig. 5d. This indicates that our experimental detuning is around 1.3 $\Omega_{m}$, which is not the optimal value due to the non-optimized laser-cavity detuning used in the experiment. Further optimization of
laser-cavity detuning would enable an optimal power threshold of 110 µW.

\begin{figure}
    \centering
    \includegraphics[width=\linewidth]{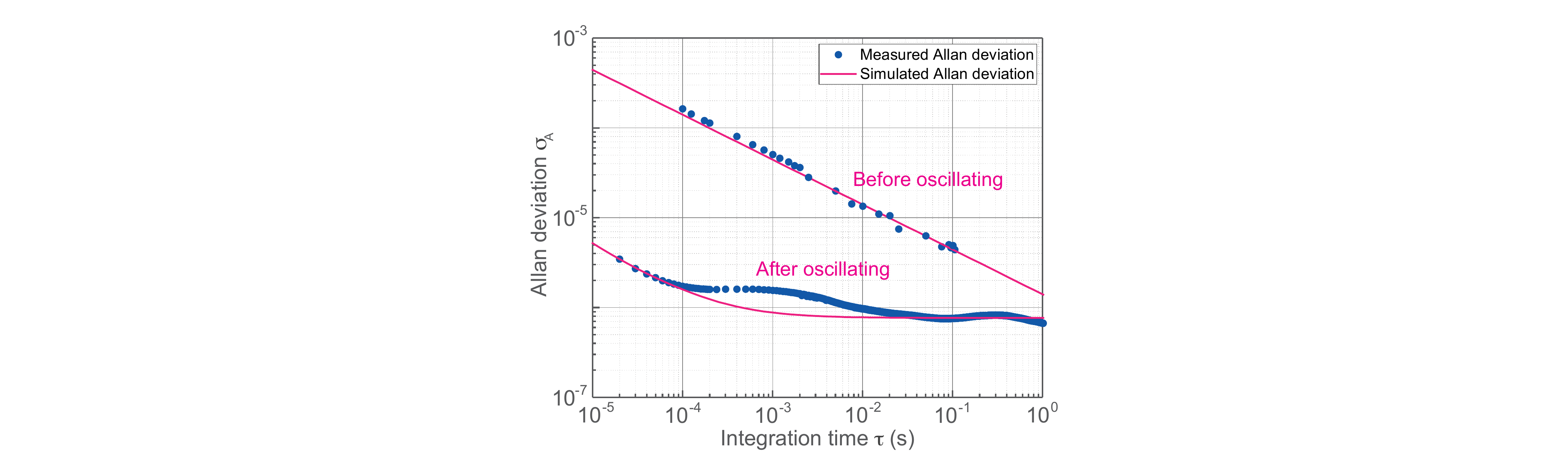}
    \caption{\textbf{Allan deviation measurement.} Measured $\sigma_A$ before and after oscillation. Blue dots denote the experimental data, while pink lines show the corresponding fits. Before oscillation, frequency instability is dominated by thermomechanical noise, yielding $\sigma_A\propto\tau^{–1/2}$. After oscillation, the overall noise reduces, but the scaling deviates at longer integration times due to the emergence of flicker noise and random-walk frequency noise.}

    \label{fig:6}
\end{figure}
Noises in an optomechanical system can limit its performance in sensing, timing, and transduction applications. We now analyze the noise sources in our optomechanical system and quantify them using NEP. Thermal fluctuation noise $\text{NEP}_T$ and thermomechanical displacement noise $\text{NEP}_\text{thermo}$ are the two fundamental noise sources on our platform and can be written as \cite{laurent201812}
\begin{equation}
    \mathrm{NEP}_T = \frac{1}{\mathrm{TCF}} \, \frac{df}{dT} 
\sqrt{\frac{4 k_B T^2 G_\mathrm{th}}{\omega^2 \tau_c^2 + 1}}~~~\text{and}~~~\mathrm{NEP}_{\mathrm{thermo}} 
= \frac{G_\mathrm{th}}{\mathrm{TCF}} \, \frac{df}{dx} 
\sqrt{\frac{k_B T}{m_{\mathrm{eff}} \, \omega_m^2 \, Q_m}},
\end{equation}
where $T$, $G_\mathrm{th}$, $\tau_c$, $\omega_m$, and $m_\text{eff}$ are the absolute temperature, thermal conductance, time constant, mechanical resonant frequency, and effective mass, respectively. Note here $\tau_c=C/G_\text{th}$, where $C$ is the thermal capacitance. With well-engineered thermal mass and heat dissipation, the device achieves a large $G_\mathrm{th}$ of 33.5 $\mu\text{W/K}$ and a small $C$ of 440 pJ/K, resulting in a $\tau_c$ of 13 $\mu$s and a thermal bandwidth of 12.2 kHz. Further decreasing $G_\mathrm{th}$ can reduce both $\text{NEP}_T$ and NEP$_\text{thermo}$, while increasing TCF and $Q_m$ can reduce NEP$_\text{thermo}$ (also Supplementary Sec.5). Thanks to the low-loss and highly isolated nature of the suspended LN resonator, optical readout noise is minimum compared with the former two noises and not considered.

We characterize the frequency stability of the device using Allan deviation ($\sigma_A$), which is subsequently converted to NEP via the device responsivity. Figure 6 compares the measured $\sigma_A$ versus integration time $\tau$ before and after oscillation. Before oscillation, due to the relatively low signal intensity ($\sim-48$ dBm), we use an oscilloscope to capture the output signal and extract the Allan deviation (see Methods). In this case, the upper limit of measurable $\tau$ is constrained by the memory depth of the oscilloscope. Within the measured range, $\sigma_A$ scales as $\tau^{–1/2}$, indicating that thermomechanical noise dominates the frequency instability. A $\sigma_A$ of $1.6 \times10^{-4}$ is obtained at an integration time of $10^{-4}$ s, corresponding to a NEP of 610 $\text{nW}/ \sqrt{\text{Hz}}$. After oscillation, the signal intensity increases significantly, allowing direct detection with a commercial frequency counter. At short integration time, thermomechanical noise remains the primary contributor, yielding a $\sigma_A$ of $1.6 \times 10^{–6}$ at $\tau=10^{–4}$ s, corresponding to a NEP of 6.2 $\text{nW}/\sqrt{\text{Hz}}$. At longer timescales, flicker noise and random walk noise become dominant (see Supplementary Fig. S4), which might originate from laser instability at high optical power and can be suppressed by adding balanced homodyne detection or PDH locking. 
Operating the device under vacuum can further reduce the thermal conductance by three orders, substantially lowering the noise floor. Besides, optimizing the anchor structure or introducing the feedback loop \cite{anetsberger2008ultralow,yu2022observation,feng2008self, rossi2018measurement} may further enhance sensitivity and suppress the noise (also see Supplementary Sec.4 and 5).

\section*{Conclusion}
We have demonstrated a room-temperature low-noise optomechanical platform on TFLN by synergistically combining an ultralow-loss optical cavity with an ultrathin membrane acting as a high-Q mechanical cavity. We demonstrate that using such highly isolated, ultralow mass resonators enables direct heat transduction to the frequency shifts of the mechanical mode with a thermal conductance and capacitance of 33.5 $\mu\text{W/K}$ and 440 pJ/K, along with a TCF of $-124$ ppm/K, exceeding those of other integrated platforms such as silicon, silicon nitride, silicon carbide, and aluminum nitride ($-20$ to $-50$ ppm/K \cite{sui2025review}). The device also features all optical interrogation with input and output optical interfaces seamlessly integrated on the same chip and harnesses the power of near-IR optical sources and photodetection technology for the shot noise-limited measurement, vastly mitigating the electronic read-out noise in conventional thermal sensors. The absence of electrical signal routing further suppresses stray noise and crosstalk, enabling robust, scalable operation in large-format sensor arrays. The combination of low thermomechanical noise and strong environmental isolation yields a NEP of 6.2 $\text{nW}/\sqrt{\text{Hz}}$ at 10 kHz. These properties can be further improved through meandered tether geometries \cite{perello2024geometrically} and acoustic bandgap engineering \cite{maccabe2020nano}, which reduce the effective thermal mass and push the thermal conductance toward the radiative limit, enabling ultrafast response and enhanced thermal sensitivity. Benefiting from the strong piezoelectric response of LN, an external optoelectronic feedback loop can be implemented via electro-mechanical actuation \cite{han201510} to suppress phase noise, analogous to techniques previously demonstrated in optoelectronic oscillators \cite{chembo2019optoelectronic}. More importantly, optical-detuning-sensitive self-oscillation enabled by such feedback provides a pathway to significantly enhance the effective TCF through the thermo-optic effect \cite{horvath2023sub}. Operation under reduced pressure offers an additional route to noise suppression: at an ambient pressure of 1 mbar, $G_{th}$ is expected to decrease by three orders of magnitude, leading to a projected NEP as low as 6.2 $\text{pW}/\sqrt{\text{Hz}}$. Owing to its compact footprint, scalability, and material versatility, this platform can be readily scaled up and integrated with tunable mid-infrared absorbers to realize highly sensitive, low-noise multispectral detectors with low cost, size, weight, and power (C-SWaP). These capabilities position the platform as a promising building block for large-scale optomechanical systems for applications such as hyperspectral remote sensing, thermal imaging, and multifunctional on-chip sensing.

\section*{Methods}
\textbf{Fabrication.} The TFLN samples under test are fabricated on an X-cut lithium-niobate-on-insulator wafer from NanoLN with a 300-nm-thick top LN layer and a 2-µm-thick bottom silicon oxide layer on a silicon substrate. Microring resonators and bus waveguides are first patterned in positive resist (PMMA) with electron-beam lithography (EBL) and transferred to the LN using angled argon milling, leaving an 80-nm-thick slab. By optimizing the etching recipe, we significantly reduced the trenching effect, enabling the realization of thinner and more uniform slabs. After stripping the resist, a thorough acid clean is performed to remove re-deposited amorphous LN. A second EBL defines the anchors and air-holes in PMMA, followed by fully etching using $\text{Ar}^+$-based inductively coupled plasma reactive ion etching (ICP-RIE). After repeating the cleaning process, the chips are annealed at 500 °C for 3 hours to restore the LN material quality. Finally, the optomechanical resonators are released by wet etching in 49\% hydrofluoric acid and dried in a critical point dryer.

\textbf{Measurement.} A tunable continuous-wave laser (Santec TSL-570) is coupled into on-chip edge couplers via a lensed fiber and collected by a collimator, with a total coupling loss of 18.8 dB, mainly due to facet roughness caused by hand cleaving. The device under test (DUT) is mounted on a commercial TEC for temperature-dependent measurement. A homodyne setup is used to measure the phase shift of the output light induced by mechanical vibrations while simultaneously suppressing laser noise. The input laser is split into two paths: one coupled into the resonator to bring out the mechanical signal, and the other serving as a phase-stable local oscillator. Interference between the two paths converts the phase modulation into an intensity signal, which is detected by the balanced photodetectors, enhancing the signal-to-noise ratio. Meanwhile, Pound-Drever-Hall locking is used to lock the laser at the resonant wavelength, which has a maximum phase response slope and thus allows the most sensitive phase readout. A signal analyzer is used to directly capture PSDs of mechanical modes, and an oscilloscope is used to monitor mechanical frequency stability in the time domain.

Here, we use an oscilloscope to measure the mechanical signal in a long time domain and derive the Allan deviation of our mechanical signal. Note to resolve 10 kHz frequency change of an 83.7 MHz signal, the time window of measurement should be more than 1 ms with a sampling rate larger than 200 MS/s. Figure S5a plots the output signal in 1 s detected by a high-speed photodetector and captured by an oscilloscope. The recorded signal is then divided into equal time slices of 1 ms, and a Fourier transform is applied to each segment to extract the instantaneous mechanical frequency in the time domain, as shown in Fig. S5b. We can then derive the Allan deviation at different integration times using the above equation, as shown in Fig. S5c.

\begin{acknowledgments}
This work is funded by DARPA under the Optomechanical Thermal Imaging (OpTIm) program (HR00112320022). M.Y. and Y.Y. are supported by the U.S. Department of Energy, Office of Science, Basic Energy Sciences, Materials Sciences and Engineering Division under Contract No. DE-AC02-05CH11231 within the Quantum Coherent Systems Program KCAS26. I.A. is supported by NASA Space Technology Graduate Research Opportunity (NSTGRO) grant No. 80NSSC24K1369. Device fabrication was performed at the John O’Brien Nanofabrication Laboratory at University of Southern California. The views, opinions and/or findings expressed are those of the authors and should not be interpreted as representing the official views or policies of the Department of Defense or the U.S. Government.
\end{acknowledgments}

\bibliography{ms.bib}

\end{document}